\documentclass[reprint,prd,nofootinbib]{revtex4-1}
\usepackage{graphicx}
\usepackage{xcolor}
\usepackage{amsmath}
\usepackage[abbreviations]{siunitx}
\usepackage{subfigure}
\usepackage[hidelinks]{hyperref}
\def\equationautorefname#1#2\null{%
  Eq.\;(#2\null)%
}
\def\figureautorefname#1\null{%
  Fig.#1\null
}

\begin{document}

\title{Electrically charged strange  stars with an interacting quark matter equation of state}
\author{V. P. Gon\c{c}alves}
\affiliation{ High and Medium Energy Group, Instituto de F\'{\i}sica e Matem\'atica,  Universidade Federal de Pelotas (UFPel)\\
Caixa Postal 354,  96010-900, Pelotas, RS, Brazil.
}
\author{L. Lazzari}
\affiliation{ High and Medium Energy Group, Instituto de F\'{\i}sica e Matem\'atica,  Universidade Federal de Pelotas (UFPel)\\
Caixa Postal 354,  96010-900, Pelotas, RS, Brazil.
}

\begin{abstract}
The properties of electrically charged strange quark stars predicted by an interacting quark matter equation of state (EoS) based on cold and dense perturbative quantum chromodynamics (pQCD) are investigated. The stability of strange stars is analyzed considering different models for the electric charge distribution  inside the star as well as for distinct values for the total electric charge. A comparison with the predictions derived using the MIT bag model is also presented.
We show that the presence of a net electric charge  inside strange stars implies in a larger maximum mass in comparison to their neutral counterparts. Moreover, we demonstrate that the pQCD EoS implies larger values for the maximum mass of charged strange stars, with very heavy charged stars being stable systems against radial oscillations. For an  electric charge distribution   given by $q(r) = \beta r^3$, the pQCD EoS implies unstable configurations for large values of the renormalization scale as well as for large values of $\beta$, in contrast to the MIT bag model predictions. 

\end{abstract}



\maketitle

\section{Introduction}
\label{sec:intro}

The description of matter at high densities and/or high temperatures  is one of the main challenges of the strong interactions theory -- the Quantum Chromodynamics (QCD) (For a recent review see, e.g. Ref. \cite{review_qcd}). While the regime of high temperatures and vanishing baryon density have been explored in heavy ion collisions at RHIC and LHC, 
the QCD at high baryon densities and low temperatures is fundamental to determine the properties of  compact stars \cite{Baym:2017whm}, where the density of the matter is predicted to exceed the nuclear density matter and the system is expected to be  described in terms of deconfined quark degrees of freedom.
According to the  the Bodmer-Witten hypothesis~\cite{bodmer1971,witten1984}, the absolute ground state for the hadronic matter is the strange quark matter (SQM) rather than $^{56}$Fe, which implies the possible existence of compact stars entirely made of deconfined up, down and strange quarks, usually denoted strange quark stars (SQS)~\cite{alcock1986,alcock1988,glendenning1996,weber1999}. 
{Another possibility is the presence of deconfined quark matter inside neutron stars (NSs), forming hybrid stars. In this case, SQM is expected to be formed through nucleation processes, converting a previously hadronic star (HS) into a SQS with the liberation of a large amount of energy that generates a neutrino burst and an intense gravitational waves emission. In fact, both HSs and SQSs can exist in Nature, in the so called two-families scenario, where neutron stars with masses up to \num{1.5}\,M$_\odot$ are HSs and the those with mass above this threshold are SQSs~\cite{drago2016}. These distinct scenarios have been investigated in detail by several authors over the last years. Its predictions compared with recent experimental data constrain the equation of state (EoS) that describes matter in NSs (for recent reviews see, e.g. \cite{Baym:2017whm,bombaci2016,drago2016}). Robust astrophysical constraints on the EoS came from the two solar mass limit of the pulsars PSRJ1614 - 2230 and PSR J0348 + 0432~\cite{pul1,pul2,pul3} and from the LIGO/Virgo detection of gravitational waves originating from the NS merger event GW170817~\cite{GW170817}. In particular, the electromagnetic and gravitational wave information from the GW170817 event  have been used to constrain the radii $R$, maximum mass $M_{max}$ and tidal properties of NSs.} {In addition, a recent result~\cite{nature} pointed out that dense matter in the interior of massive NSs ($M \approx 2 M_{\odot}$) exhibits characteristics of a deconfined quark phase in the core of the star. Even so, the interpretation of these recent results still is a theme of intense debate and the existence of hybrid and/or strange quark stars remains an open question.}




As pointed out before, the discovery of pulsars with large masses ($M \gtrsim 2$ M$_\odot$)~\cite{cromartie2019}  has  put  strong constraints  on  the  equation  of  state  (EoS)  of  dense stellar matter and has challenged the description of these  objects as being  quark stars, which were predicted to have smaller masses by models based on the phenomenological MIT bag model EoS. However,  the main properties of these compact objects strongly depend on a precise description of the matter that composes the star \cite{weber1999}. In particular, it is the EoS that defines the magnitude of the internal pressure that competes with gravity and, consequently, that establishes the stability of the star. During the last years, several phenomenological models have been proposed to describe the EoS for the deconfined quark system, considering different assumptions and approximations for the description of the interaction between quarks, as well as for the treatment of the running quark masses and coupling constant (See, e.g. Refs. \cite{Lugones:2002va,Ivanov:2005be,kurkela2010,fraga2014,Torres:2012xv, Zacchi:2015lwa,Kojo:2014rca,Benic:2014jia,Xu:2015wya,Chen:2016ran,Dutra:2015hxa,xia2017}). {In particular, in Ref.~\cite{kurkela2010}, the authors have derived an EoS based on cold and dense perturbative QCD (pQCD). They were able to estimate the pressure at non-zero density at order $\alpha_s^2$, where $\alpha_s$ is the strong coupling constant, assuming a non-zero value of the strange quark mass.} One advantage of this EoS, is that it allows to estimate the systematic uncertainty present in a perturbative calculation. The results presented in Refs.~\cite{kurkela2010,fraga2014} demonstrated that strange quark stars with masses larger than 2 M$_\odot$ can be reached for large values of the renormalization scale.

In addition to the EoS, the maximum mass of the star can be modified by the presence of electric charge in its interior. {Neutron stars, so as SQSs, are electrically neutral. However, to achieve electrical neutrality in quark matter one must allow the possibility of leptons being present. In particular, for the latter, the presence of electrons is necessary to form a chemically equilibrated system, where  $\beta$-equilibrium is achieved. Such electrons are distributed in a layer on the surface of the star, which is separated from strange matter by several hundred Fermi. These two systems interact through electrostatic force, resulting in a charge distribution inside the star. In general, stars} can remain in equilibrium under their own gravity and electric repulsion, with the Coulombian force acting in addition to the gradient pressure to counterbalance the gravitational attraction, which implies that the charged SQS can be more stable than the neutral one. As pointed out in Refs.~\cite{usov,negreiros2009,arbanil2015}, the surface of a SQS has a high electric field, which can reach about \SI{1e21}{V/m}. The results presented in~\cite{negreiros2009,arbanil2015} and~\cite{rincon} using the massless version of the MIT bag model and a non-linear EoS, respectively, have demonstrated that charged stars are heavier than their neutral counterparts, with the increasing in the maximum  mass being dependent of the magnitude of the electric charge. {As a consequence, the presence of the electric charge can modify the star compactness, given by the ratio between its mass and radii ($M/R$), which determines tidal deformability. Such property have been determined using the recent data for the GW170817 event, which have also been used to constrain the EoS (See e.g. Refs. \cite{Sieniawska:2018zzj,Christian:2018jyd,Wang:2019npj}).}
   
Our goal in this paper is to present, for the first time, the predictions from pQCD 
for charged SQSs. In particular, we will investigate the hydrostatic equilibrium considering the interacting quark matter EoS derived in Refs.~\cite{kurkela2010,fraga2014} and different models for the electric charge distribution inside the star, as well as for different values of the total charge. In addition, we will investigate the impact of radial oscillations on the stability of charged SQS. Our study is  strongly motivated by the possibility to use its predictions and experimental tests to improve our understanding of the inner structure of compact objects as well as to constrain the EoS of the system (See e.g. Ref. \cite{kokkotas2001}). Previous calculations for radial modes in neutral and charged SQS were performed in Refs.~\cite{VasquezFlores:2010eq,brillante2014,jimenez2019,negreiros2009,arbanil2015} considering different models for the EoS. {Our aim is to investigate the dependence on the EoS of the dynamical stability of charged SQS against radial perturbations considering different assumptions for the electric charge  distribution.}

This paper is organized as follows. In the next section, we will present a brief review of the formalism used to describe charged SQSs and {their radial oscillations}. 
The different models used to describe the charge distribution in the star will be discussed and the pQCD EoS presented. In~\autoref{sec:res}, we present our results for the mass-radius profile and for the fundamental mode of oscillation considering the pQCD EoS and distinct models for the treatment of charge inside the star. The predictions derived using the MIT bag model considering massive quarks are also presented, which improve the analysis performed in Ref.~\cite{arbanil2015} and allow a detailed comparison with the pQCD results. Finally, in~\autoref{sec:conc} we summarize our main conclusions. In what follows, we use the units $c = 1 = G$.

\section{Formalism}
\label{sec:theory}

The presence of  charge in a SQS implies that the system should be described by the Einstein-Maxwell field equations, { with the energy density associated to the  electric field being present into the energy - momentum tensor. One has that due to the high value of the electric field on the surface of the star, the electric energy density is of the same order as the energy density of the strange quark matter. As a consequence, the high electric field expected on the surface of a SQS, affects the space-time metric and the associated energy density contributes to its own gravitational mass.} 
 Moreover, the presence of the Coulomb interaction modifies the structure equations that describe the relativistic hydrostatic equilibrium (See Ref.~\cite{brillante2014}). 
{In order to describe a spherically symmetric static charged star, we will assume a line element given by}
\begin{equation}
  \label{eq:ds2}
  \mathrm{d}s^2 = e^{2\nu(r)}\mathrm{d}t^2 - e^{2\lambda(r)}\mathrm{d}r^2 - r^2(\mathrm{d}\theta^2 + \sin^2\theta\, \mathrm{d}\phi^2)\,,
\end{equation}
where $\lambda$ and $\nu$ are known as the metric functions. Such metric implies that a charged star constituted by a perfect fluid will satisfy the following stellar structure equations 
\begin{align}
  \label{eq:TOV-q}
  \frac{\mathrm{d}q}{\mathrm{d}r} & {} = 4\pi r^2 \rho_e e^{\lambda} \,,\\
  \label{eq:TOV-m}
  \frac{\mathrm{d}m}{\mathrm{d}r} & {} = 4\pi r^2 \epsilon  + \frac{q}{r}\frac{\mathrm{d}q}{\mathrm{d}r} \,, \\
  \label{eq:TOV-p}
  \frac{\mathrm{d}P}{\mathrm{d}r} & {} = -(\epsilon + P)\left(4\pi r P + \frac{m}{r^2} - \frac{q^2}{r^3}\right)e^{2\lambda}   \nonumber \\
  & {} \quad + \frac{q}{4\pi r^4}\frac{\mathrm{d}q}{\mathrm{d}r} \,, \\
  \label{eq:TOV-nu}
  \frac{\mathrm{d}\nu}{\mathrm{d}r} & {} = -\frac{1}{\epsilon + P}\left(\frac{\mathrm{d}P}{\mathrm{d}r} - \frac{q}{4\pi r^4}\frac{\mathrm{d}q}{\mathrm{d}r}\right) \,,
\end{align}
where $\rho_e(r)$ is the electric charge density, $q(r)$ and $m(r)$ represent the charge and mass within  radius $r$, respectively. The metric potential $ e^{-2\lambda} $ has the Reisser-Nordstr\"om form
\begin{equation}
  \label{eq:e2lambda}
  e^{-2\lambda(r)} = 1 - \frac{2m(r)}{r} + \frac{q(r)^2}{r^2}\,.
\end{equation}
For neutral stars one has that $q(r) = 0$ and~\autoref{eq:TOV-p} reduces to the usual Tolman-Oppenheimer-Volkoff equation. 
To solve such a system of equations, one needs to establish the boundary conditions. At the center of the star we assume that
\begin{equation}
  \label{eq:r0}
  q(0) = m(0) = 0\,,\qquad \epsilon(0) = \epsilon_{\mathrm{c}}\,,\qquad \nu(0) = \nu_{\mathrm{c}}\,.
\end{equation}
Moreover, we assume that the solutions on the surface of the star ($r=R$) satisfy the following conditions
\begin{align}
p(R) & {} = 0\,, \\
m(R) & {} = M\,, \\
q(R) & {} = Q\,, \\
\nu(R) & {} = -\lambda(R)\,,
\end{align}
{where $M$ and $Q$ are the total mass and electric charge of the stellar system, respectively. }

In order to investigate the stability of the charged SQS against radial oscillations, we will consider the approach proposed by Chandrasekhar~\cite{chandrasekhar1964} many years ago, which demonstrated that perturbing the fluid and space-time variables in a manner that maintains the spherical symmetry of the system, it is possible to derive an equation for infinitesimal radial oscillations of a spherical object: the pulsation equation. Such equation is given by
\begin{equation}
  \label{eq:stumliouville}
  \frac{\mathrm{d}}{\mathrm{d}r}\left[{\cal{P}} \frac{\mathrm{d}u}{\mathrm{d} r} \right] + [{\cal{Q}} + \omega^2{\cal{W}}]u = 0
\end{equation}
where $u$ is the renormalized displacement function and for a charged star we have that~\cite{brillante2014}
\begin{align}
  \label{eq:pqw}
  \cal{P} & {} = e^{\lambda+3\nu}r^{-2}\gamma P\,, \\
  \cal{Q} & {} = (\epsilon + P)r^{-2}e^{\lambda + 3\nu}\left[\nu'\left(\nu' - 4r^{-1}\right) - (8\pi P  + r^{-4}q^2)e^{2\lambda}\right] \,,\\
  \cal{W} & {} = e^{3\lambda+\nu}r^{-2}(\epsilon + P)\,,
\end{align}
where $\gamma$ is the adiabatic index.
The pulsation equation constitutes a Sturm-Liouville eigenvalue problem, which allows to obtain the eigenvalues and eigenfunctions of the radial perturbation. Defining the auxiliary variable $\eta \equiv {\cal{P}}{\mathrm{d}u}/{\mathrm{d}r}$ we can transform the above second-order differential equation into two first order differential equations given by 
\begin{equation}
  \label{eq:eta}
  \frac{\mathrm{d}u}{\mathrm{d}r} = \frac{ \eta}{\cal{P}} \,, 
\end{equation}
and
\begin{equation}
  \label{eq:detadr}
  \frac{\mathrm{d} \eta}{\mathrm{d} r} = -[{\cal{Q}} + \omega^2{\cal{W}}]u \,.
\end{equation}
In Ref.~\cite{kokkotas2001}, the authors demonstrated that for $\eta(0) = 1$ one has $u(0) = r^3/(3{\cal{P}}(0))$, which are the initial conditions for the integration of the equations from the origin to the surface of the star. As in Refs.~\cite{arbanil2015,kokkotas2001} we will use the shooting method to obtain the values of $\omega^2$ that satisfies the boundary condition given by
\begin{equation}
  \label{eq:dudr}
  \frac{\mathrm{d}u}{\mathrm{d}r}\bigg\vert_{r=R} = \eta(R) = 0\,.
\end{equation}

The structure equations are solved using a Runge-Kutta-Cash-Karp method with adaptive step size. Starting from a trial value for $\omega^2$, we obtained the values that satisfy the boundary conditions using the Newton-Raphson method. This values are the eigenfrequencies of the pulsation equation. 
One important aspect is that for charged SQS the condition $\partial M/\partial \epsilon_{\mathrm{c}} > 0$ is not sufficient to determine the stability of the star~\cite{arbanil2015}. Therefore, in order to investigate the stability of such objects, one has to perform the analysis of its radial perturbation modes. Since ${\cal{Q}}$ is real, the next eigenfrequency is always larger than the previous one, i.e.,
$$\omega_0^2 < \omega_1^2 < \omega_2^2 \cdots < \omega_n^2 < \cdots \,.$$ Consequently, to determine the stability of the star it is sufficient to analyze the sign of the fundamental mode. For $\omega_0^2 < 0$ the star is unstable. 


In order to solve the structure equations, we must specify the charge distribution and the EoS that describes the matter inside the star. In our analysis, motivated by the studies performed in Refs.~\cite{arbanil2015,deb}, we will consider the following models to describe the charge in the SQS:
\begin{itemize}
\item Model A: The charge is proportional to the third power of the radial coordinate as follows
\begin{equation}
  \label{eq:qbeta}
  q(r) = Q\left(\frac{r}{R}\right)^3 \equiv \beta r^3\,,
\end{equation}
where $\beta \equiv Q/R^3$;
\item Model B: The charge density is proportional to the energy density, i.e.,
\begin{equation}
  \label{eq:qalpha}
  \rho_e = \alpha\epsilon\,,
\end{equation}
where, in geometric units, $\alpha$ is a dimensionless proportionality constant;
\item Model C: The star has a fixed total charge $Q$.
\vspace{0.5cm}
\end{itemize}
{The main difference between models A and B is associated to the region inside the star where the charge is expected to be larger. In model A, the charge is almost totally concentrated on the surface of the star. While in model B there is a relevant amount of charge in the intermediate region between the center of the star and its surface. In contrast, model C is independent of the model assumed for the electric charge distribution  inside the star.}   

Regarding the EoS, the simplest model and more frequently used to describe the interior of a quark star is the MIT bag model~\cite{chodos1974}, which characterizes a degenerate Fermi gas of up, down and strange quarks. In such model, the main properties  only depend on the bag constant $B$. However, the MIT bag model is a naive approximation, which is not sufficiently powerful to characterize a system with interacting quarks or more complex structures. In our analysis, we will consider the pQCD EoS calculated in Ref.~\cite{kurkela2010} at order $\alpha_s^2$ and a non-zero value of the strange quark mass. This description was put in a simple to use formula in Ref.~\cite{fraga2014}, being given by 
\begin{equation}
  \label{eq:pQCD-EoS}
  P = P_{\mathrm{SB}}(\mu_B)\left(c_1 - \frac{a(X)}{\mu_B - b(X)}\right)\,,
\end{equation}
where
\begin{equation}
  \label{eq:psb}
  P_{\mathrm{SB}}(\mu_B) = \frac{3}{4\pi^2}\left(\frac{\mu_B}{3}\right)^4\,
\end{equation}
corresponds to the pressure of a gas composed by three massless non-interacting quarks, also called a Stephan-Boltzmann (SB) gas, and the functions $a(X)$ and $b(X)$ are auxiliary functions (for details, see Ref.~\cite{fraga2014}). The dimensionless parameter $X$ is proportional to the renormalization scale parameter $\bar{\Lambda}$ that arises in the perturbative expansion and is expressed as $X = 3\bar{\Lambda}/\mu_B$. Fixing $X$, the energy density comes from the following relation
\begin{equation}
  \label{eq:eQCD}
  \epsilon = -P + \mu_B n_B\,,
\end{equation}
where $n_B$ is the baryon number density obtained from the thermodynamical relation $$n_B = \frac{\partial P}{\partial \mu_B}\,.$$
In the next section, we will present our predictions for the mass-radius profile of the SQS as well as for the fundamental mode considering the  pQCD EoS and the distinct models for the distribution of charge discussed above. 

\begin{figure*}[t!]
  \begin{tabular}{ccc}
    \includegraphics[width=.34\textwidth]{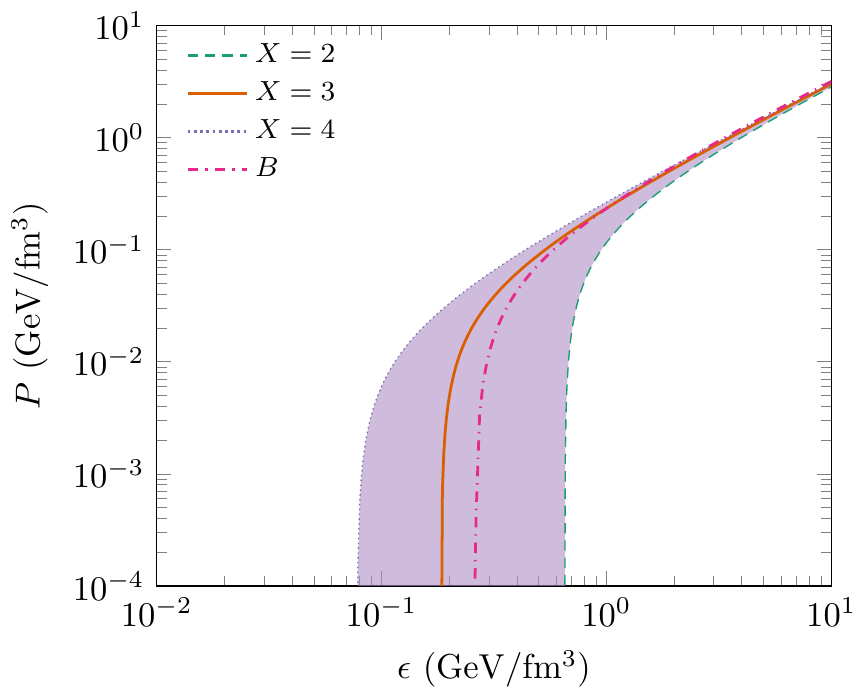} &  \includegraphics[width=.32\textwidth]{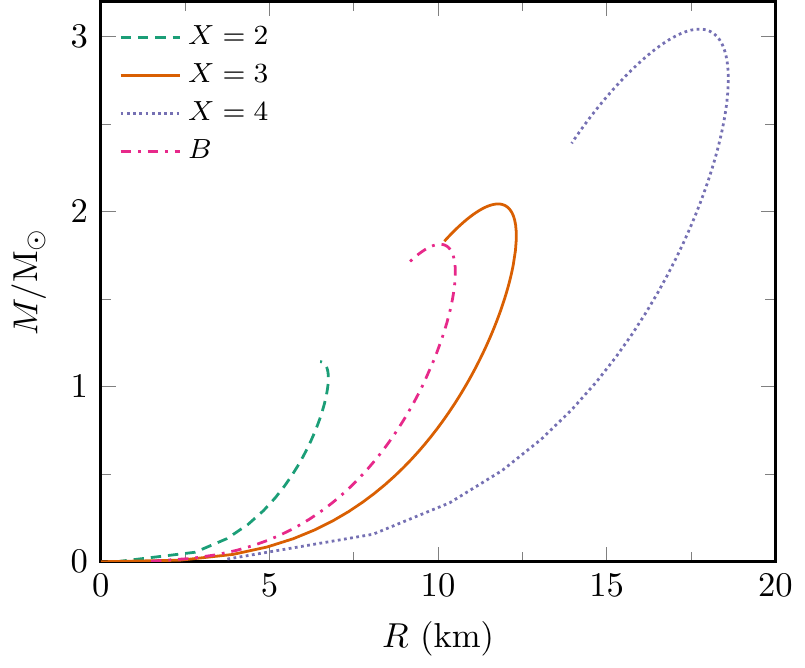} &  \includegraphics[width=.31\textwidth]{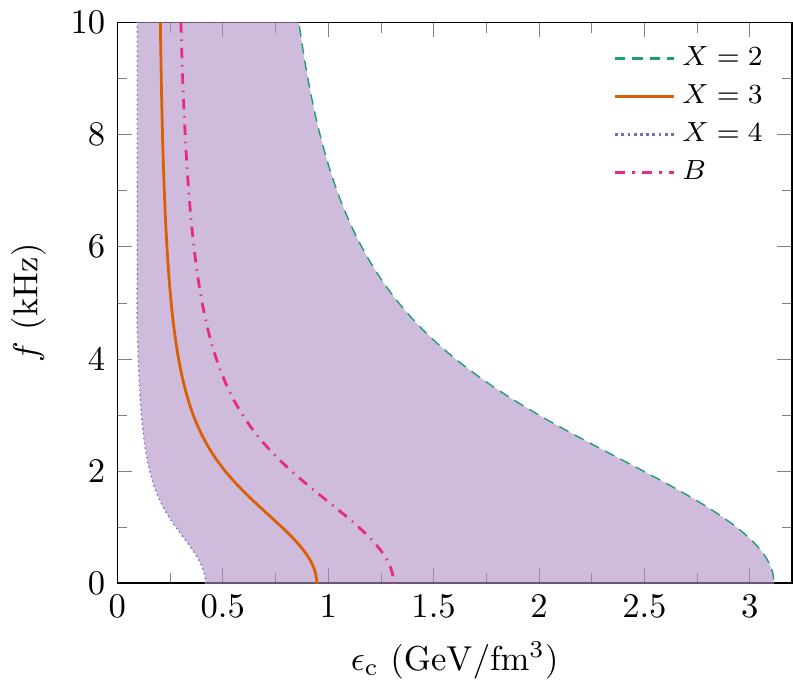} \\ 
    (a) & (b) & (c)
  \end{tabular}
  \caption{(a) Comparison between the MIT bag model EoS and the pQCD one considering different values of the renormalization scale $X$; (b) Mass -- radius profile and (c) linear fundamental frequency for a neutral SQS derived assuming the MIT bag model and pQCD EoS's.}
\label{fig:neutral}
\end{figure*}

\begin{figure*}[!t]
  \centering
  \subfigure{\includegraphics[width=.35\textwidth]{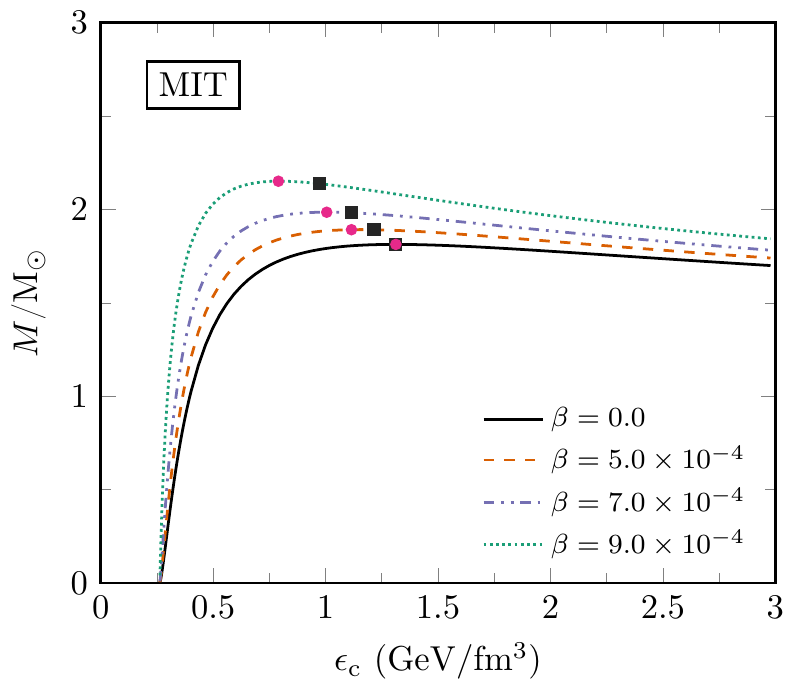}}\quad
  \subfigure{\includegraphics[width=.35\textwidth]{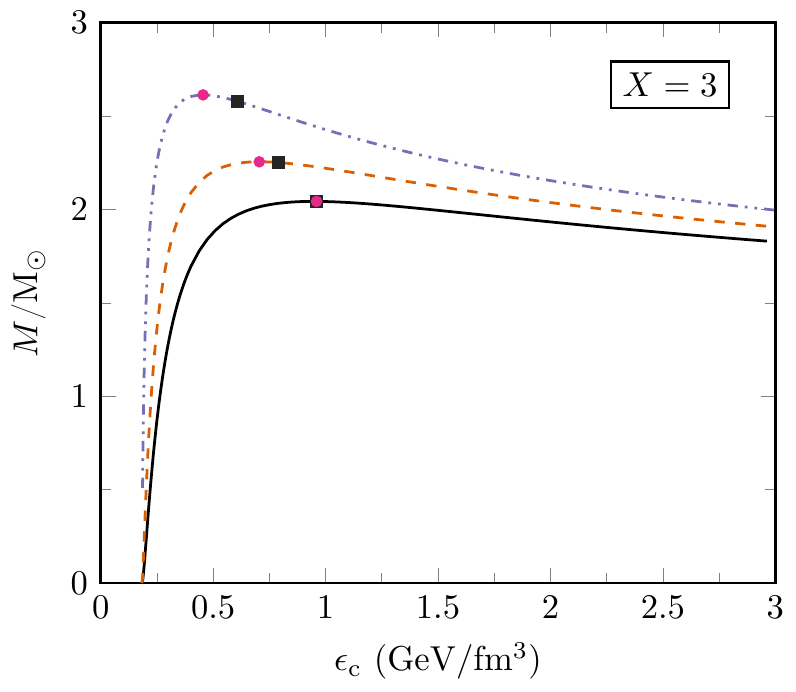}}

  \subfigure{\includegraphics[width=.35\textwidth]{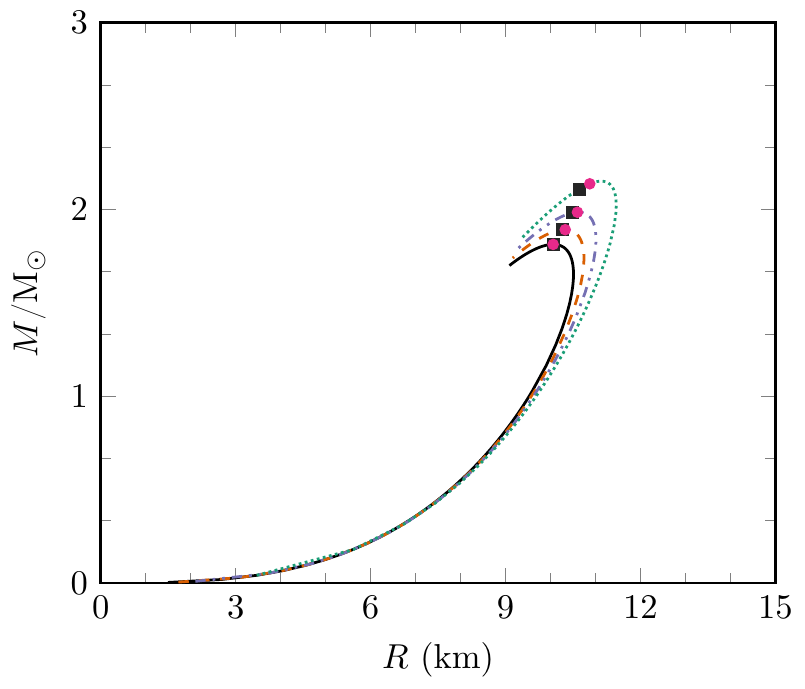}}\quad
  \subfigure{\includegraphics[width=.35\textwidth]{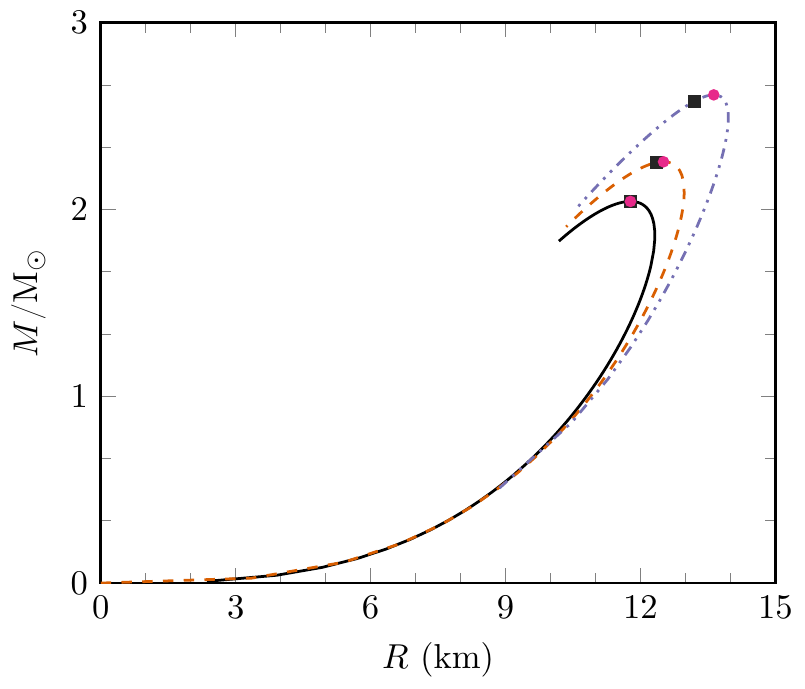}}
  
   \subfigure{\includegraphics[width=0.35\textwidth]{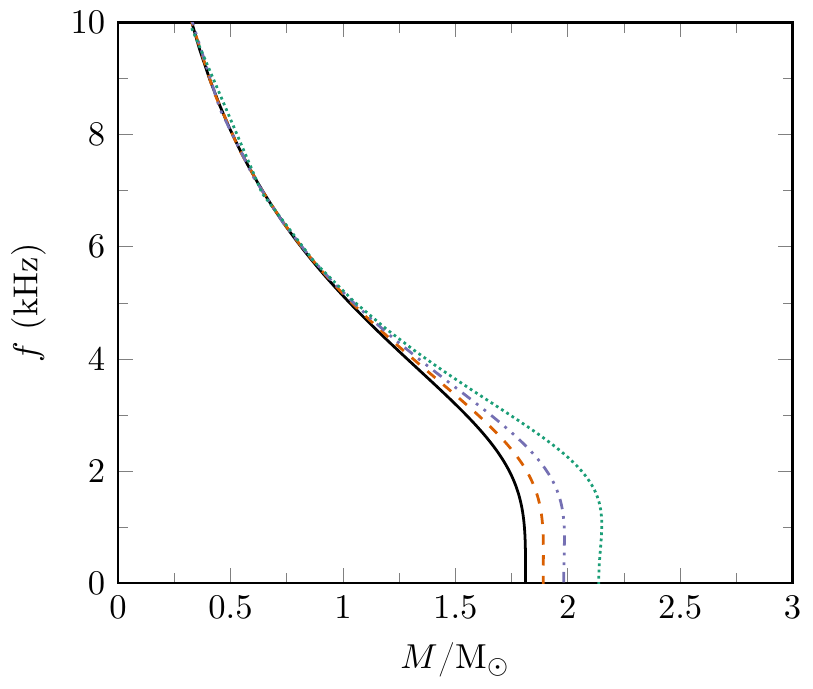}}\quad
  \subfigure{\includegraphics[width=0.35\textwidth]{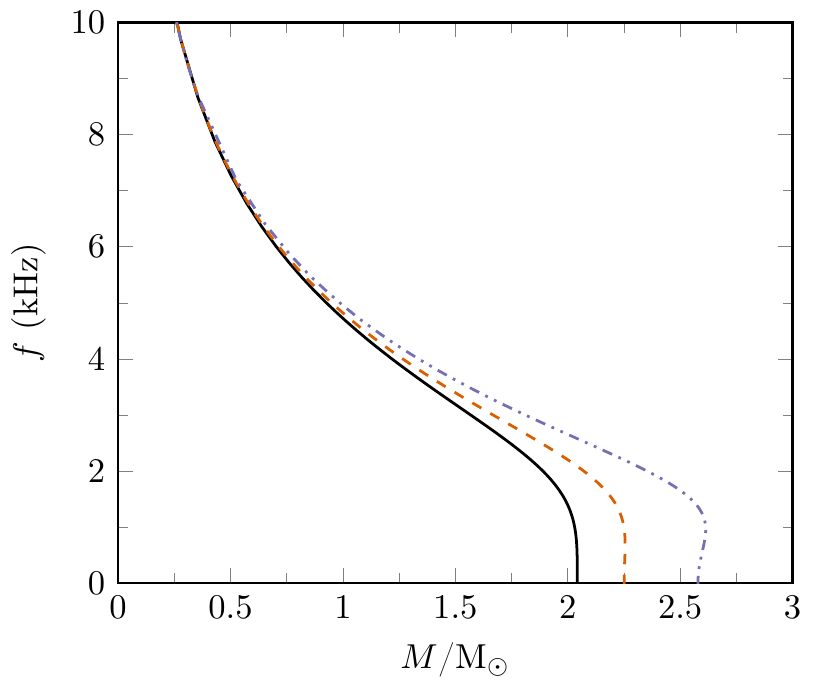}}
  
  \caption{Predictions from the MIT bag model (left panels) and pQCD (right panels) EoSs for the total gravitational mass as a function of the  central energy density ({upper panels}) and   radius ({central panels}), as well as for the linear fundamental frequency as a function of the mass (lower panels) considering  Model A and different values of $\beta$, where $\beta = 0$ (black solid line) corresponds to a neutral strange star. The full squares represent the configurations for which $\omega_0^2 = 0$ and the full circles represent the maximum mass configurations.}
  \label{fig:beta-pQCDxMIT}
\end{figure*}

\section{Results}
\label{sec:res}

In~\autoref{fig:neutral} (a), we present a comparison between the EoSs for the quark matter predicted by the MIT bag model and the by the cold and dense pQCD calculation performed in Ref.~\cite{kurkela2010}. For comparison, we present the  MIT bag model predictions derived assuming that the bag pressure is \SI{60}{MeV.fm^{-3}} and that the strange quark has a mass of $m_s = \SI{150}{MeV}$. For the pQCD EoS we present the predictions derived assuming different values for the renormalization scale $X$. We have that the pQCD EoS is strongly dependent on the renormalization scale, with the band representing the uncertainty associated to this scale.

For completeness, let us initially solve the structure equations for neutral SQS, in which $q(r) = 0$, for the distinct EoSs discussed above. Our results for the mass-radius profile  are presented in ~\autoref{fig:neutral}(b). We can see that the SQSs maximum masses depend strongly on the value of $X$, increasing with $X$ and reaching values larger than 2 M$_\odot$  for $X \gtrsim 3$~\cite{fraga2014}. We have verified that the MIT bag model predictions are similar to those derived using the pQCD EoS for $X \approx 2.8$, which predicts values of maximum SQS masses smaller than 2 M$_\odot$. As already pointed out in Ref.~\cite{jimenez2019}, only values of $X$ in the range between 3 and 3.2 satisfy simultaneously the GW170817 constraints of mass and radius~\cite{GW170817}. {The studies performed in  Refs. ~\cite{margalit2017,shibata2017,rezzolla2018} suggested that  the upper limit on the neutron star mass is 2.17 M$_{\odot}$ (90\% credibility). }


We present in~\autoref{fig:neutral} (c) the comparison between the fundamental eigenfrequencies obtained from pQCD (for different values of $X$) and the MIT bag model as a function of the central energy density. For convenience, we  are presenting results for the linear frequency associated to the eigenfrequency by $f = \omega/2\pi$. The results indicate that the configurations for the distinct values of $X$ are stable against radial oscillations ($\omega_0^2 > 0$), in agreement with the results presented in Ref.~\cite{jimenez2019}.




\begin{figure*}[!t]
  \centering
  \subfigure{\includegraphics[width=.35\textwidth]{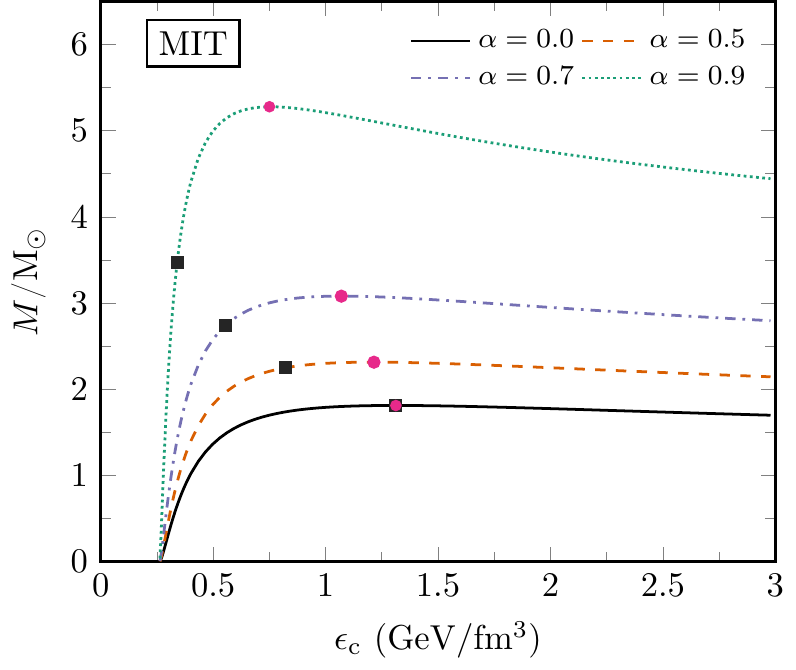}}\quad
  \subfigure{\includegraphics[width=.35\textwidth]{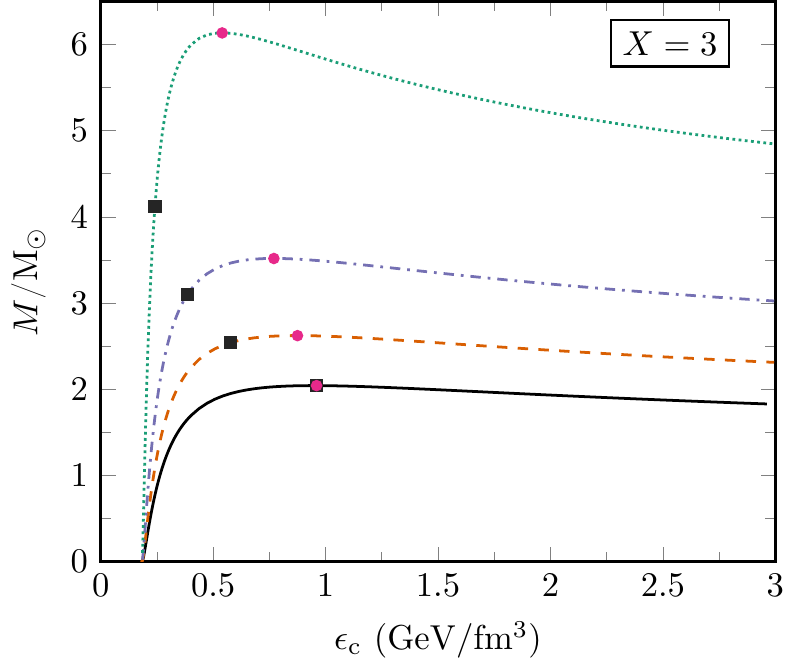}}

  \subfigure{\includegraphics[width=.35\textwidth]{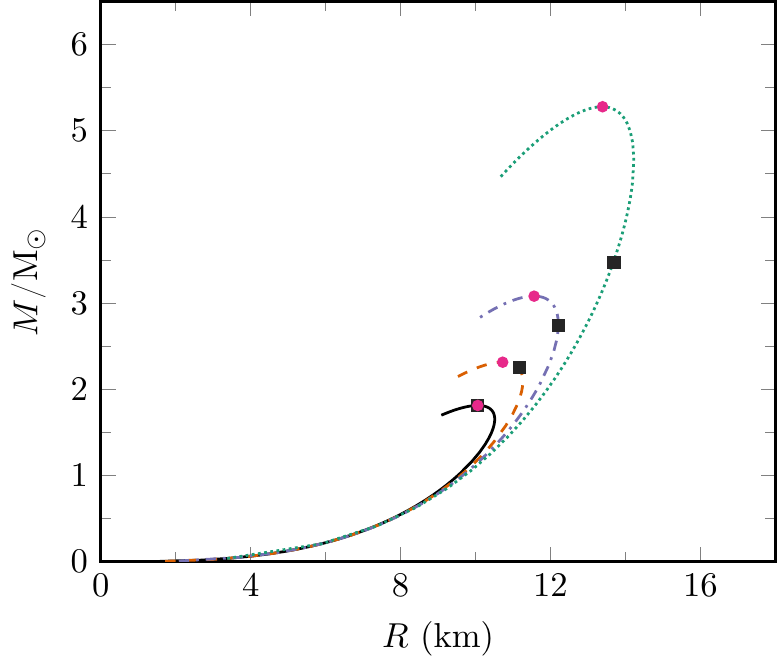}}\quad
  \subfigure{\includegraphics[width=.35\textwidth]{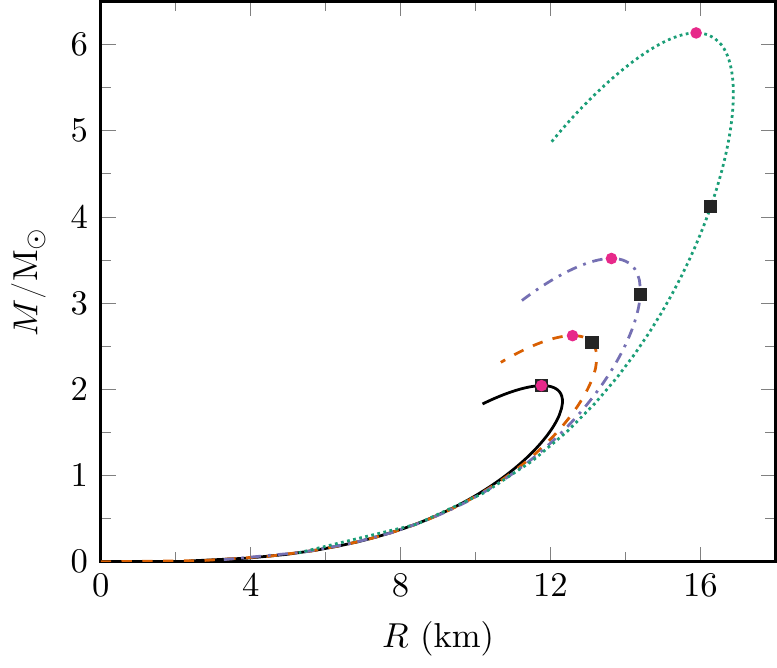}}
  
   \subfigure{\includegraphics[width=.35\textwidth]{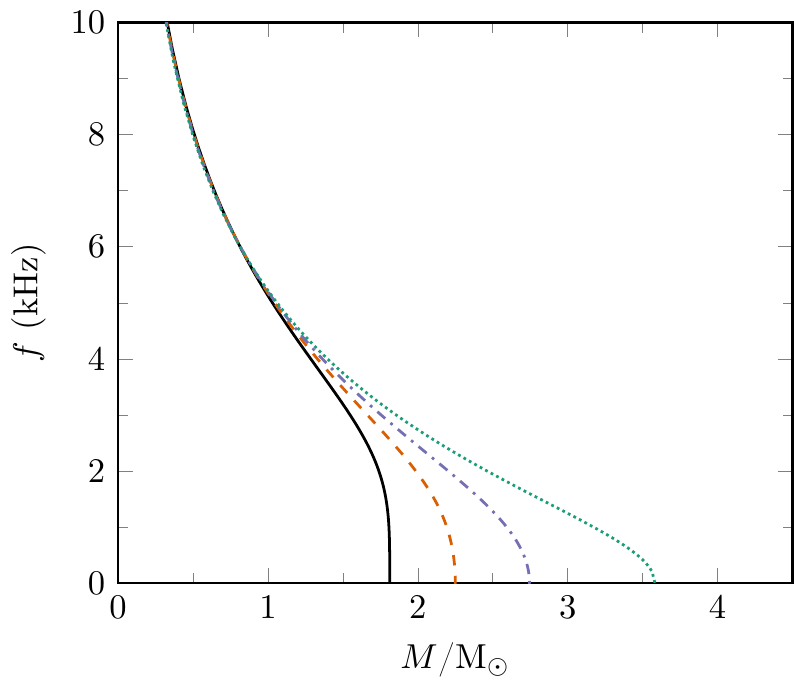}}\quad
  \subfigure{\includegraphics[width=.35\textwidth]{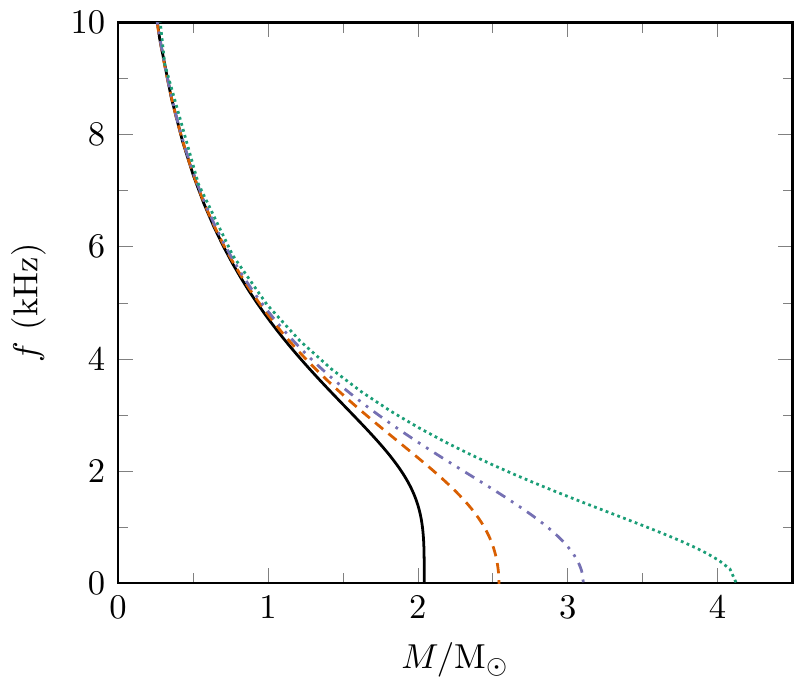}}
  
  \caption{Predictions from the MIT bag model (left panels) and pQCD (right panels) EoSs for the total gravitational mass as a function of the  central energy density ({upper panels}) and   radius ({central panels}), as well as for the linear fundamental frequency as a function of the mass (lower panels) considering  Model B and different values of $\alpha$, where $\alpha = 0$ (black solid line) corresponds to a neutral strange star. The full squares represent configurations for which $\omega_0^2 = 0$ and the full circles represent the maximum mass configurations.}
  \label{fig:alpha-pQCDxMIT}
\end{figure*}

In what follows, we will present our results for charged SQS considering the pQCD EoS for $X=3$, which satisfies the GW170817 constraints, { and the distinct models for the electric charge distribution discussed above}.
 For comparison, we will also present the predictions derived using the massive MIT bag model EoS with $B = \SI{60}{MeV.fm^{-3}}$ for the bag pressure. First, we present in Fig.~\ref{fig:beta-pQCDxMIT} our predictions for model A, in which the charge distribution charge is given by $q(r) =  \beta r^3$, assuming different values of $\beta$,  {with $\beta = 0$ corresponding to the neutral SQS.}
Our results show that for both EoSs the presence of charge increases the maximum mass and radius of the star in comparison to their neutral counterpart. Such result agrees with those derived in Ref.~\cite{negreiros2009}, which pointed out that increasing the  charge allows the star to sustain larger masses. However, the predictions derived using the pQCD EoS are even more sensitive to the presence of charge in the star, since the same variation of $\beta$ leads to a considerable larger modification in the mass and radius of the respective stellar configurations in comparison to the MIT bag model predictions. The full circles and squares in the figures indicate, respectively, the points where the maximum mass configuration occurs and where the fundamental eigenfrequency is zero. For $\beta = 0$ we have that these two points coincide. Therefore, for neutral SQS, $\partial M/\partial \epsilon_{\mathrm{c}} > 0$ is a necessary and sufficient condition to determine the configurations of stable equilibrium~\cite{arbanil2015}. In contrast, for charged SQS, these two points are not coincident, with the zero eigenfrequency configuration occurring for larger central densities. Consequently, in order to determine the stability of a charged SQS the signal of the fundamental mode  should also be analyzed~\cite{arbanil2015}.
{In our analysis, using the MIT bag model EoS, we have obtained that increasing $\beta$ up to $\num{9e-4}\,$M$_\odot\si{.km^{-3}}$ always produces stable configurations in which $\partial M/\partial \epsilon_{\mathrm{c}} > 0$ and $\omega_0^2 > 0$. In contrast, for the pQCD EoS, the largest value of $\beta$ considered, $\beta = \num{9e-4}\,$M$_\odot\si{.km^{-3}}$, gives an unstable solution for all values of $\epsilon_{\mathrm{c}}$. This explains why the associated predictions is not presented on the right panels of Fig.~\ref{fig:beta-pQCDxMIT}. It is important to emphasize that we have verified that a similar instability occurs in the MIT bag model predictions for $\beta \gtrsim \num{5e-3}\,\mathrm{M}_\odot\si{.km^{-3}}$. }

In Fig. \ref{fig:alpha-pQCDxMIT}, we present our predictions using model B, in which the charge density is proportional to the energy density $\rho_e = \alpha\epsilon$, considering different values for the dimensionless constant $\alpha$.
Similarly to the results obtained for model A, the charge density proportional to the energy density also implies in SQSs with larger masses and radii. Again, the impact of increasing the values of $\alpha$ is larger on the pQCD EoS. The main difference between models A and B is associated to the fact that
model B is stable for all values of $\alpha$ under analysis, independently of the EoS being considered. In particular, for large values of $\alpha$, model B predicts stable configurations with very high masses and radii. 
{This difference occurs because, in model B, the changes in  mass and  charge depend on the energy density, which implies that the increasing of the electric field  throughout the star is exactly counterbalanced by  gravity. In contrast, in model A, the increasing of the electric field is very fast near to the surface of the star. At large values of the electric charge, the outward pressure associated to the electric field cannot be sustained by   gravity, and the associated configuration is unstable.  Even though stable charged SQSs with very large masses and radii are predicted by model B, it is important to emphasize that the largest mass pulsar ever observed has a mass of {  2.14$^{+0.10}_{-0.09}$ M$_\odot$ for the 68.3\% credibility interval and 2.14$^{+0.20}_{-0.18}$ M$_\odot$ for the 95.4\% credibility interval ~\cite{cromartie2019}}. Therefore,  charged SQSs with highest mass values predicted by  model B, although mathematically stable,  probably do not exist in Nature.}


\begin{figure*}[!t]
  \centering
  \subfigure{\includegraphics[width=.35\textwidth]{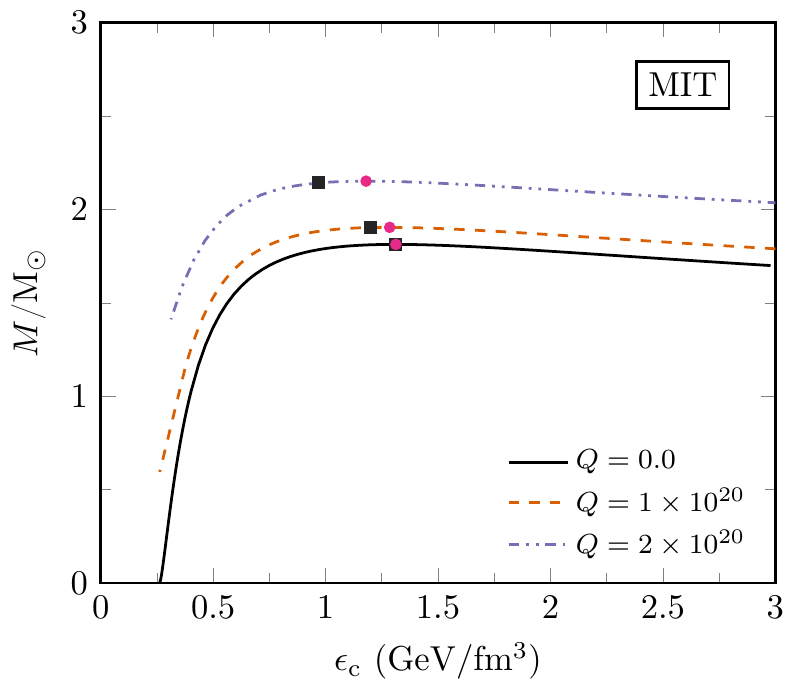}}\quad
  \subfigure{\includegraphics[width=.35\textwidth]{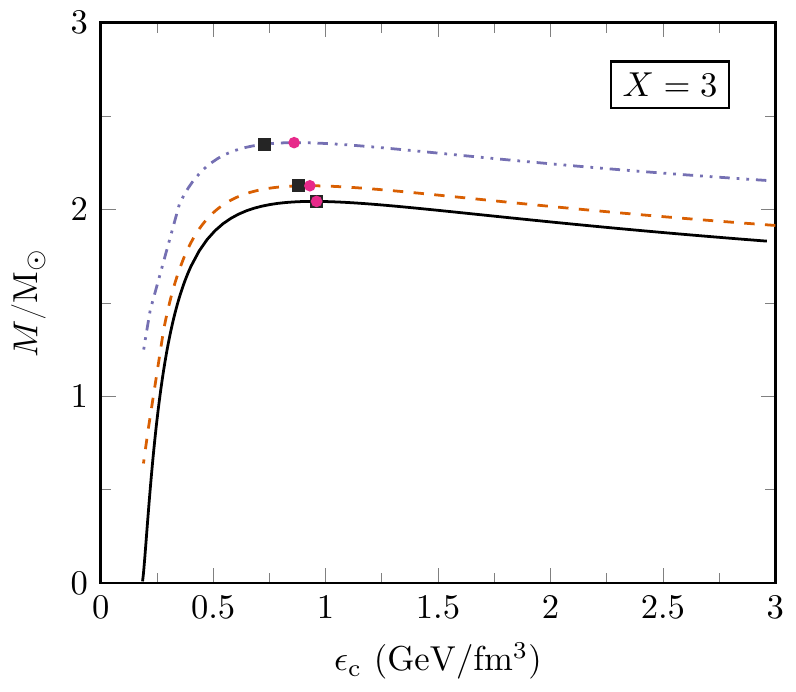}}
  
  \subfigure{\includegraphics[width=.35\textwidth]{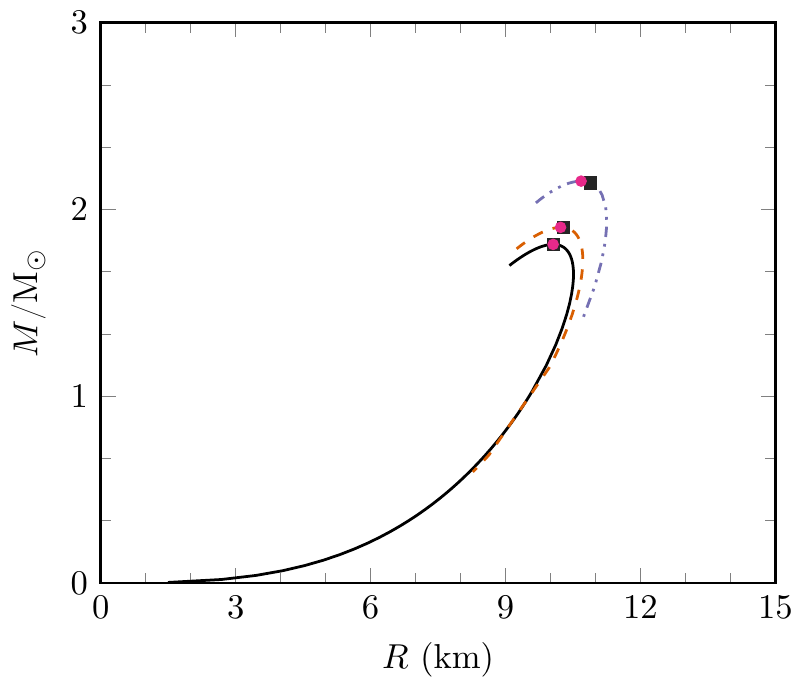}}\quad
  \subfigure{\includegraphics[width=.35\textwidth]{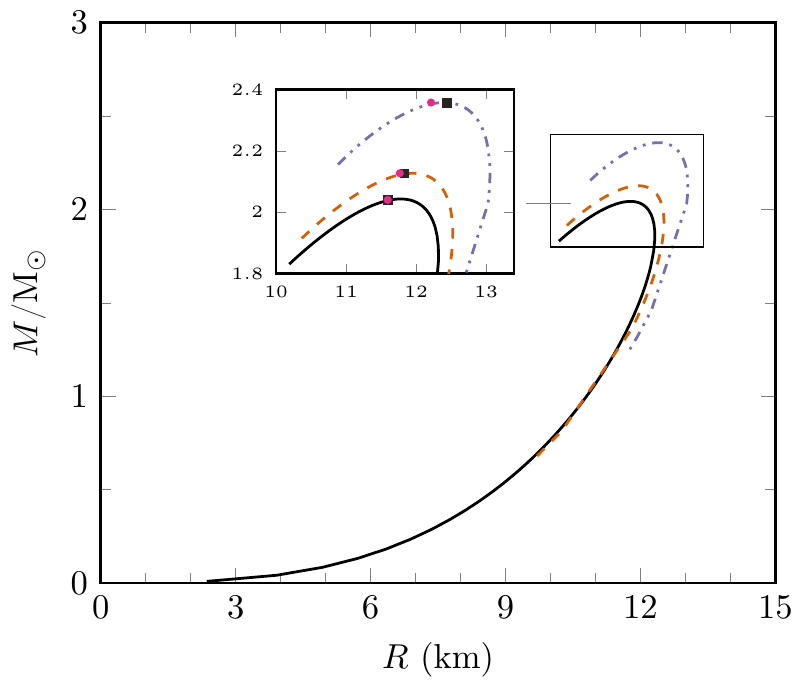}}
  
   \subfigure{\includegraphics[width=.35\textwidth]{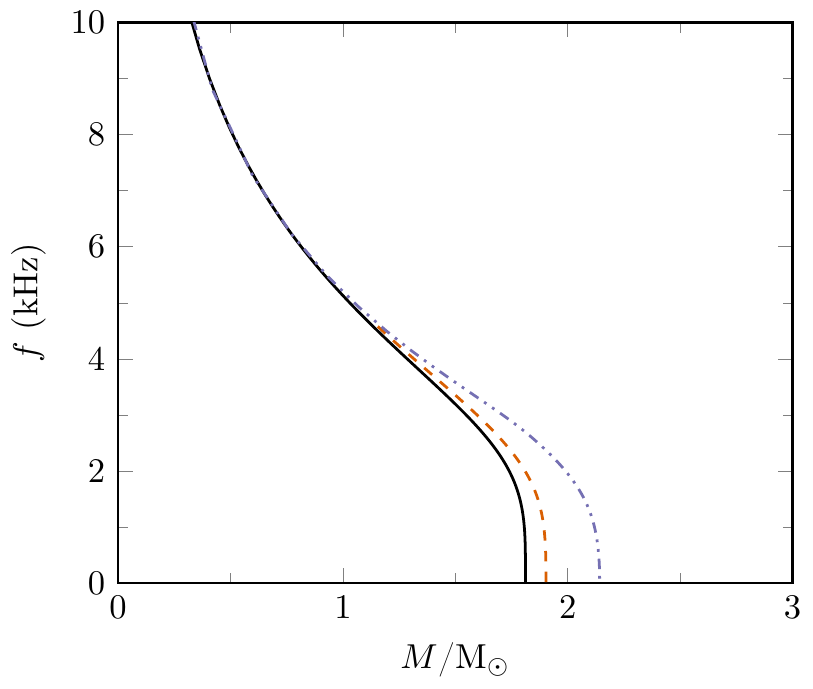}}\quad
  \subfigure{\includegraphics[width=.35\textwidth]{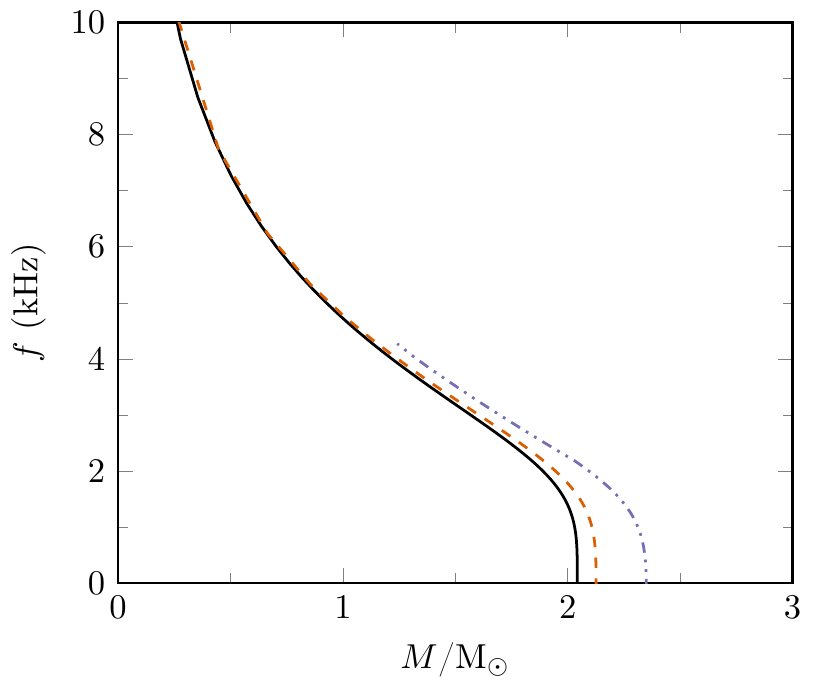}}
  
  \caption{Predictions from the MIT bag model (left panels) and pQCD (right panels) EoSs for the total gravitational mass as a function of the  central energy density ({upper panels}) and  the radius ({central panels}), as well as for the linear fundamental frequency as a function of the mass (lower panels) considering  Model C and different values of the total charge $Q$, where $Q = 0$ (black solid line) corresponds to a neutral strange star. The full squares represent the configurations for which $\omega_0^2 = 0$ and the full circles represent the maximum mass configurations.}
  \label{fig:fixedQ}
\end{figure*}

Finally, let's consider model C, in which we assume a star with a fixed total charge $Q$. From Gauss' law for electric fields, we have that such assumption implies that the predictions are independent of the charge distribution inside the star. Our results are presented in Fig.~\ref{fig:fixedQ} for different values of the total charge $Q$. As in the previous models, model C also predicts that the presence of charge increases the maximum mass of the star. Our results also indicate that the point where the configuration with maximum mass occurs is slightly distinct from the position where the frequency of the fundamental mode is zero, with the distance between these points increasing with the value of the total charge $Q$, in contrast with the results obtained in Ref.~\cite{arbanil2015}.

{ Two comments are in order.  First,} in our analysis, we only have presented the predictions derived using the pQCD EoS for $X = 3$. However, we also have performed  the analysis for other values of $X$. For values of $X < 3$, the results are similar to those presented above, with the main difference being that the predicted values of the maximum masses of the neutral and charged SQS are smaller. On the other hand, for $X = 4$ and using model A for the electric charge distribution, we have found that the presence of charge implies in unstable configurations, independent of the value of $\beta$. We have verified that for  $\beta = \num{5e-4}\,\mathrm{M}_\odot\si{.km^{-3}}$, the configurations outside equilibrium start to appear when $X \geq \num{3.6}$. Such result demonstrate the strong dependence of the predictions for the charged SQS on the EoS considered. 
{Second, we have estimated the compactness ($M/R$) of the star for the different values of the charge $Q$ and verified that it increases for larger values of $Q$. The increasing is expected to modify the tidal deformability of the star and, therefore, can be used to discriminate between the neutral and charged scenarios. Such result motivates a more detailed study, which we intend to perform in the near future.}


\section{Summary}
\label{sec:conc}
In this paper we have investigated, for the first time, the equilibrium and stability of charged strange stars considering the EoS derived in Ref.~\cite{kurkela2010}, which takes into account the interaction between quarks and have been derived using cold and dense perturbative QCD. {The predictions from the pQCD EoS for SQSs are expected to be more realistic in comparison to those derive e.g. using the MIT bag model, because it takes into account quark interactions in a more systematic way.} 
We have considered three models for the treatment of charge in the SQS and have performed a detailed comparison between the predictions derived using the pQCD and MIT bag model EoSs. For both EoSs, we have verified that  the presence of a net electric charge implies in SQSs with larger maximum masses in comparison to their neutral counterparts. However, the pQCD EoS leads to larger values for the maximum mass of the charged SQS, with very heavy charged stars being stable systems against radial oscillations. In addition, our results also demonstrated that for a distribution of electric charge inside the star given by $q(r) = \beta r^3$, the pQCD EoS implies unstable configurations for large values of the renormalization scale $X$ as well as for large values of $\beta$, in contrast to the MIT bag model predictions.

\section*{Acknowledgements}
This work was partially financed by the Brazilian funding agencies CNPq, Coordena\c{c}\~ao de Aperfei\c{c}oamento de Pessoal de N\'ivel Superior  (CAPES) -- Finance Code 001,   FAPERGS and INCT-FNA (process number 464898/2014-5).


\end{document}